\documentclass{iau} 
\usepackage{graphicx}
\usepackage{hyperref}

\title[Spiral arm contrasts and galaxy properties]
      {The sequence of spiral arm classes:\\ Observational signatures of persistent spiral density waves in grand-design galaxies}

\author[A. Bittner et al.]{%
    A. Bittner,$^{1,2}$
    D. A. Gadotti,$^{1}$
    B. G. Elmegreen,$^{3}$
    E. Athanassoula,$^{4}$\\
    D. M. Elmegreen,$^{5}$
    A. Bosma$^{4}$ \and
    J. Mu\~noz-Mateos$^{6}$
}

\affiliation{%
$^1$European Southern Observatory, Karl-Schwarzschild-Str. 2, D-85748 Garching, Germany\\ email: {\tt adrian.bittner@eso.org}
\\[\affilskip]
$^2$Ludwig-Maximilians-Universit\"at, Professor-Huber-Platz 2, 80539 Munich, Germany
\\[\affilskip]
$^3$IBM Research Division, T.J. Watson Research Center, Yorktown Heights, NY 10598, USA
\\[\affilskip]
$^4$Aix Marseille Universit\'e, CNRS, LAM, Laboratoire d'Astrophysique de Marseille, France
\\[\affilskip]
$^5$Vassar College, Dept. of Physics and Astronomy, Poughkeepsie, NY 12604, USA
\\[\affilskip]
$^6$European Southern Observatory, Casilla 19001, Santiago 19, Chile
}

\pubyear{2019}
\volume{353}  
\jname{Galactic Dynamics in the Era of Large Surveys}
\editors{M. Valluri \& J. Sellwood}

\begin{document}

\maketitle

\begin{abstract}
We investigate how the properties of spiral arms relate to other fundamental galaxy properties.  To this end, we use
previously published measurements of those properties, and our own measurements of arm-interarm luminosity contrasts for
a large sample of galaxies, using $3.6\mu m$ images from the Spitzer Survey of Stellar Structure in Galaxies. Flocculent
galaxies are clearly distinguished from other spiral arm classes, especially by their lower stellar mass and surface
density. Multi-armed and grand-design galaxies are similar in most of their fundamental parameters, excluding some bar
properties and the bulge-to-total luminosity ratio.  Based on these results, we discuss dense, classical bulges as a
necessary condition for standing spiral wave modes in grand-design galaxies.  We further find a strong correlation
between bulge-to-total ratio and bar contrast, and a weaker correlation between arm and bar contrasts. 
\end{abstract}


\firstsection 


\section{Spiral arm classes}
The spiral structure in disc galaxies exhibits a great variety in its properties, such as the number of spiral arms,
their amplitudes, pitch angles and the overall level of symmetry. Based on this visual appearance, spiral galaxies are
classified in three different classes: \emph{flocculent}, \emph{multi-armed} and \emph{grand-design} galaxies. 
Spiral arms in flocculent galaxies (e.g. NGC\,2841, NGC\,7793) appear short, patchy and very irregular, suggesting that
their spiral structure is mainly caused by local gravitational instabilities of the old stellar component, gas, and the
resulting formation of new stars (see e.g. \cite[Elmegreen \& Elmegreen, 1984]{ee1984}). 
On the contrary, the strongly bi-symmetric spiral arms of grand-design galaxies (e.g NGC\,1566, M\,51) might well be
caused by density waves on global scales, as initially suggested by \cite{lindblad1959}. While density waves can be
driven by bars or satellites (\cite[Athanassoula, 1980]{lia1980}), they might as well form self-consistently by the
swing amplification mechanism (\cite[Toomre, 1981]{toomre1981}) or the density wave theory (\cite[Lin \& Shu,
1964]{linshu1964}). 
Multi-armed galaxies (e.g. NGC\,0628, NGC\,1232) share properties with both flocculent and grand-design galaxies, for
instance by showing bi-symmetric spirals in the centre which become more and more irregular at larger radii. Therefore,
these galaxies are usually considered an intermediate case (see e.g. \cite[Elmegreen \& Elmegreen, 1984,
1995]{ee1984,ee1995}).

While the different processes that might trigger spiral structure are somewhat well understood, it still remains unclear
why different mechanism dominate in different galaxies. In other words, how are the processes that cause the spiral
structure related to the fundamental properties of their host galaxies? 
In \cite{bittner2017} we perform a thorough comparison of the spiral arm strength and various fundamental galaxy
properties, including bars and disc breaks.  In the following, we summarise the main results, provide clear evidence
that distinguishes flocculent galaxies from the other spiral arm classes, and present observational signatures
consistent with long-lasting spiral density waves in grand-design galaxies.

\section{Properties of discs in spiral galaxies}

{\underline{\it Spiral Arm Contrasts}}
In order to quantify the visual classification in the spiral arm classes, we parametrize the strength of the spiral arms
by their arm-interarm intensity contrast.  To this end, we exploit a subsample of 288 galaxies obtained with the Spitzer
Survey of Stellar Structure in Galaxies (S$^4$G; \cite[Sheth et al. 2010]{sheth2010}). In particular, we use 3.6$\mu m$
observations, as these are less sensitive to dust emissions and highlight the old stellar component.  All images are
transformed into polar coordinates and the arm and interarm regions identified at various radii. By computing the ratio of
the intensities, we obtain radial contrast profiles.  The median of this profile in the radial range of the disc
component, as determined by multi-component photometric decompositions (\cite[Salo et al. 2015]{salo2015}), represents
the resulting arm-interarm contrast. 

\begin{figure}[t]
    \begin{center}
        \includegraphics[width=0.5\textwidth]{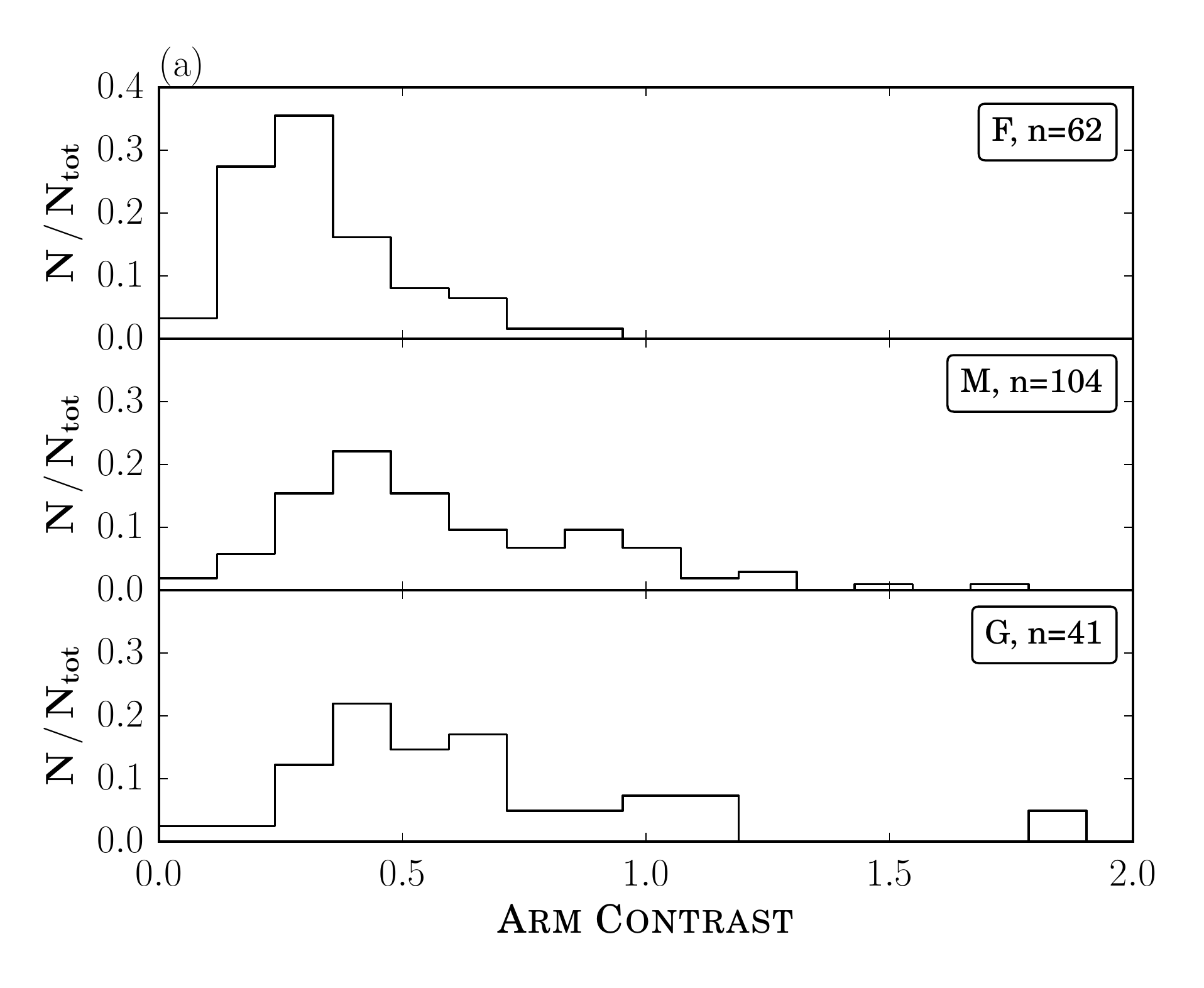}
    \end{center}
    \caption{%
        Distributions of arm contrasts for flocculent (\emph{upper panel}), multi-armed (\emph{middle panel}), and
        grand-design (\emph{lower panel}) galaxies. 
    }
    \label{fig:fig1}
\end{figure}

In Fig. \ref{fig:fig1}, we show the distributions of the spiral arm contrasts separated by spiral arm class.
Surprisingly, the increase in spiral arm strength from flocculent to grand-design galaxies is not as significant as
expected from the visual classification. In fact, only flocculent galaxies show lower spiral arm contrasts while the
distributions of multi-armed and grand-design galaxies are not significantly different. 

\begin{figure}[t]
    \begin{center}
        \includegraphics[width=0.48\textwidth]{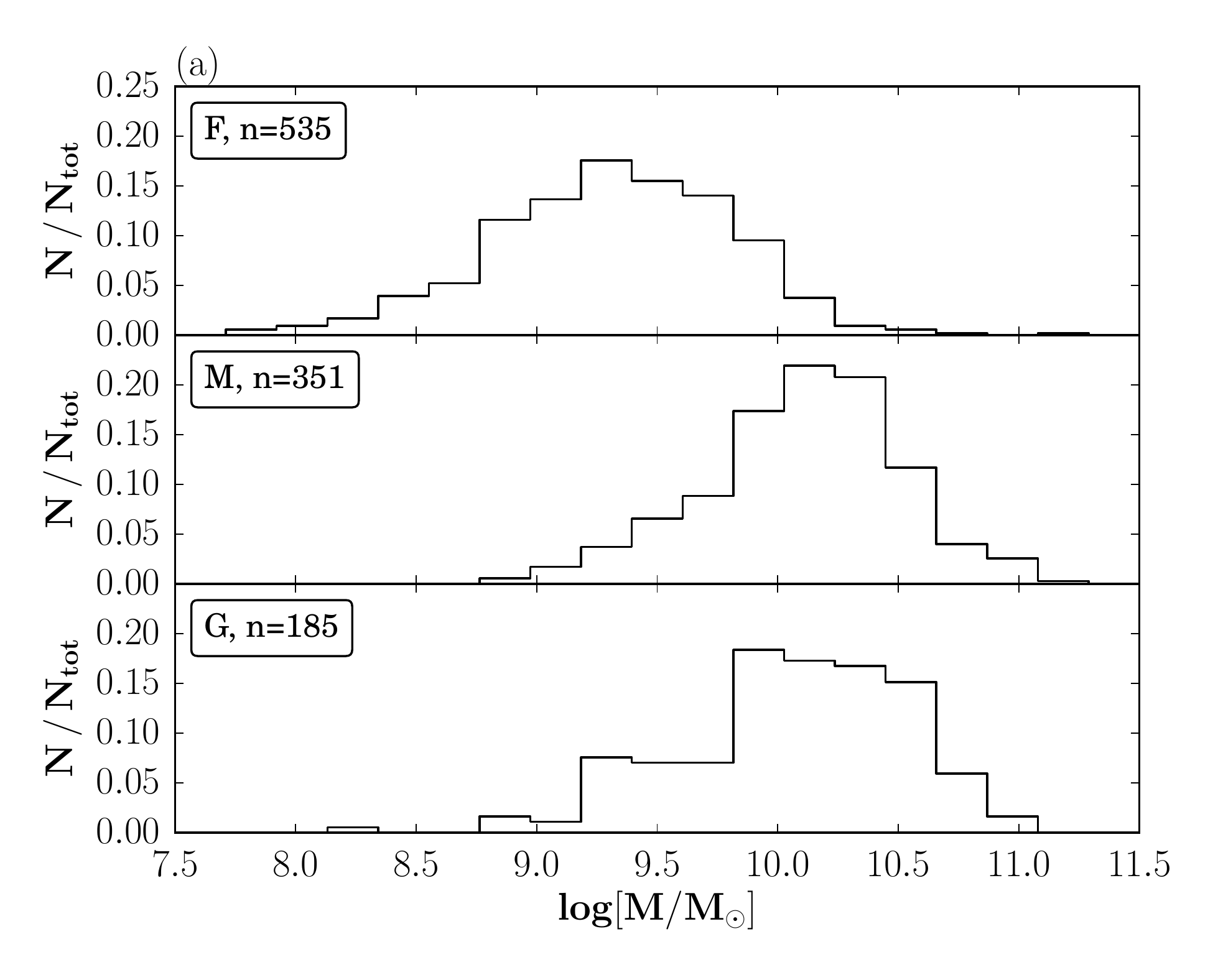}
        \includegraphics[width=0.48\textwidth]{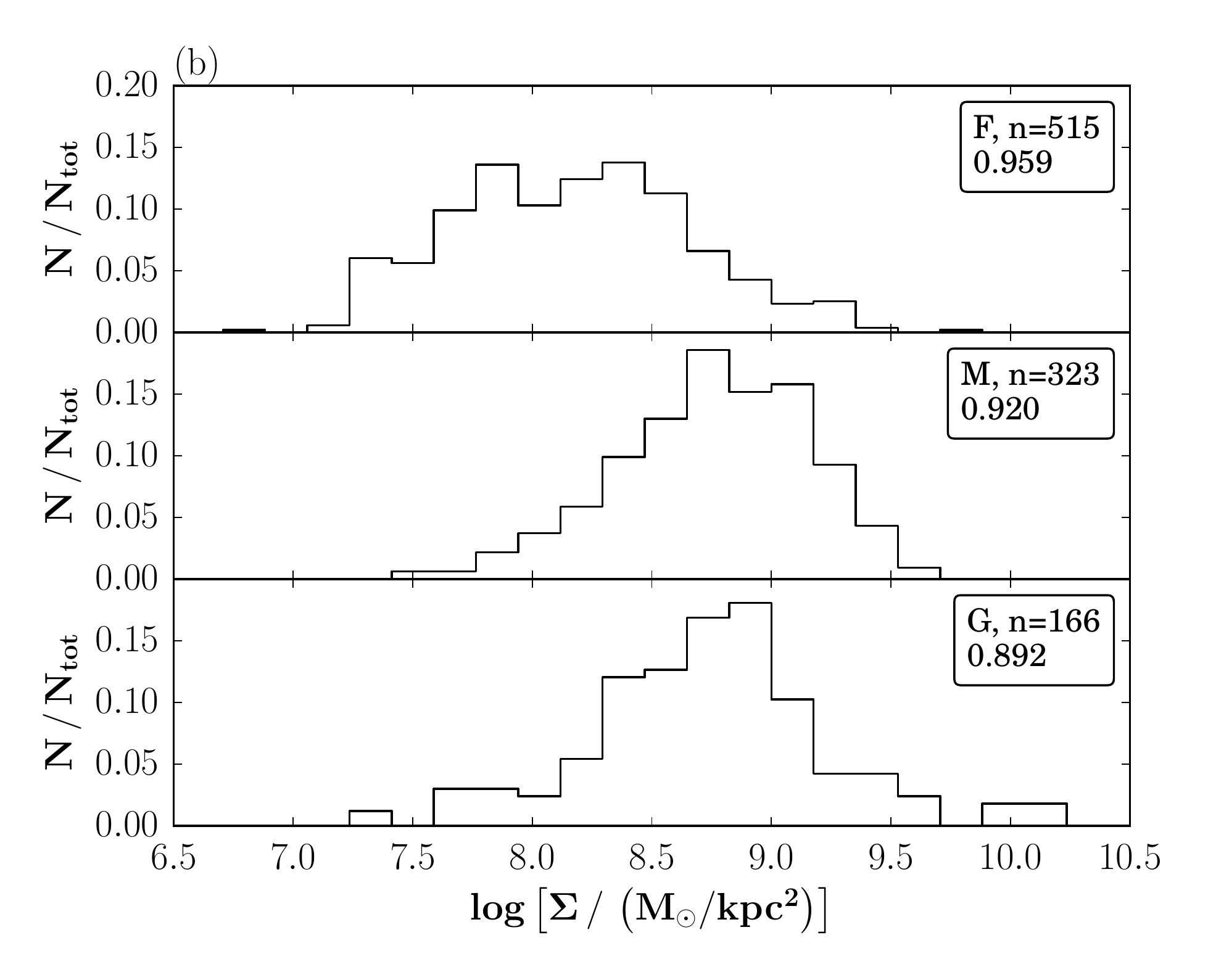}
    \end{center}
    \caption{%
        Distributions of total stellar masses (\emph{left}) and stellar mass surface densities (\emph{right}).  The
        distributions of flocculent, multi-armed, and grand-design galaxies are displayed in the \emph{upper},
        \emph{middle} and \emph{lower} panels, respectively. 
    }
    \label{fig:fig2}
\end{figure}

{\underline{\it Fundamental Galaxy Properties}}
In Fig. \ref{fig:fig2}, we present the distributions of the total stellar mass and stellar mass surface densities,
separated by spiral arm class. Here we find a similar behaviour as for the distributions of the arm-interarm contrast:
flocculent galaxies are clearly distinguished by their lower masses and surface densities, while multi-armed and
grand-design galaxies show very similar distributions. 
To further investigate these findings, we use the results of two-dimensional, multi-component decompositions, derived in
pipeline 4 of the S$^4$G survey (\cite[Salo et al., 2015]{salo2015}), as a parametrisation of the fundamental galaxy
properties. In fact, most of these considerations support the previous findings, except some bar properties and the
bulge-to-total luminosity ratio (Fig.~\ref{fig:fig3}; see also Sect.~\ref{sec:sec3} for a detailed discussion).
Interestingly, we also find a strong correlation between bulge-to-total ratio and the bar contrast, and a weak
correlation between arm and bar contrast (see Fig.~12 in \cite[Bittner et al. 2017]{bittner2017}) in multi-armed and
grand-design, but not in flocculent galaxies. 
Therefore, we conclude that the three spiral arm classes do not represent a continuous sequence, but two distinct
groups. While flocculent galaxies are hosted by a different type of discs, multi-armed and grand-design galaxies appear
to be variants of each other. 

\section{The connection between dense bulges and spiral density waves}
\label{sec:sec3}
\begin{figure}[t]
    \begin{center}
        \includegraphics[width=0.5\textwidth]{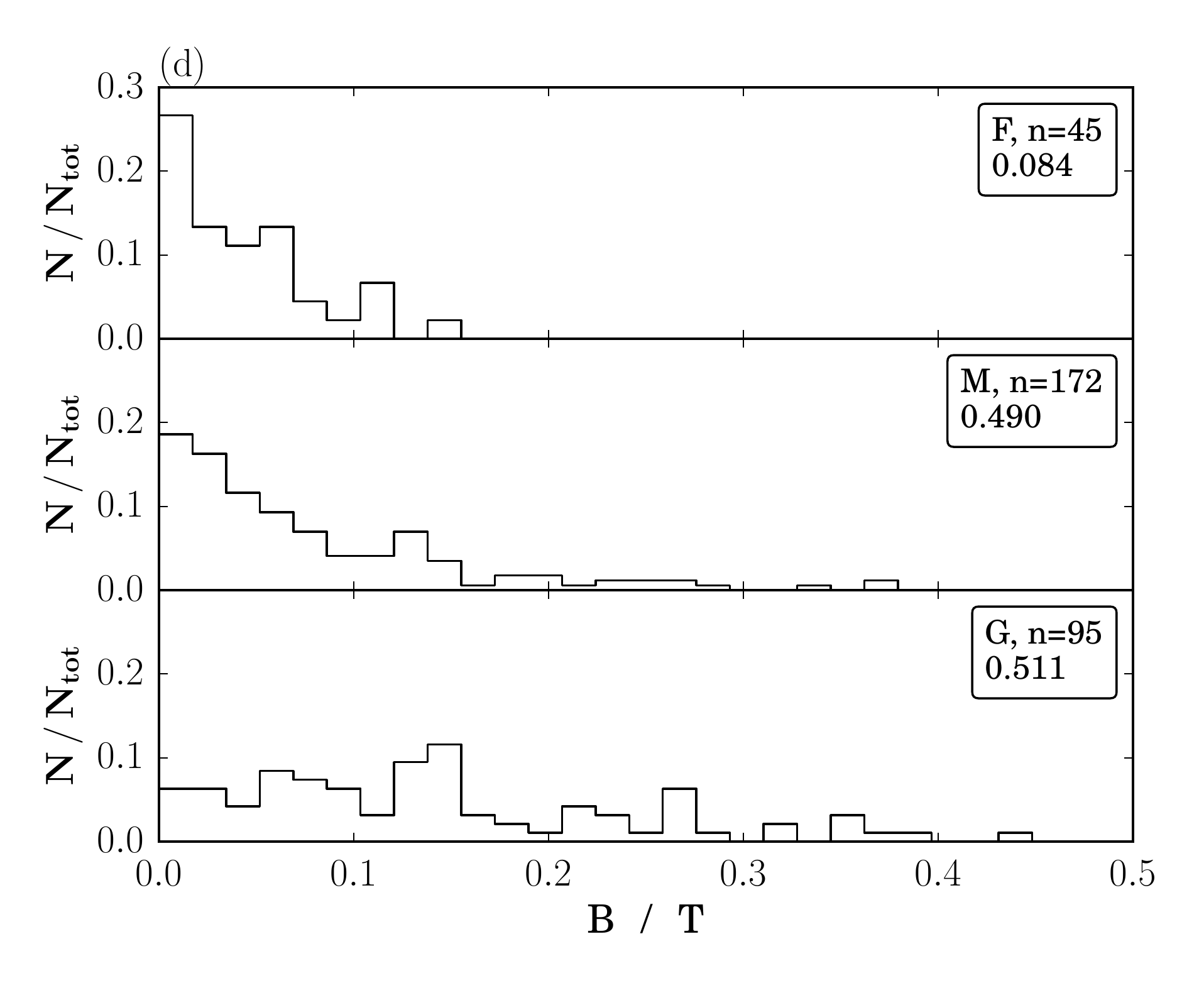}
    \end{center}
    \caption{%
        Distributions of the bulge-to-total luminosity ratio for flocculent (\emph{upper panel}), multi-armed
        (\emph{middle panel}), and grand-design (\emph{lower panel}) galaxies.
    }
    \label{fig:fig3}
\end{figure}

While multi-armed and grand-design galaxies appear vastly similar in most of their fundamental galaxy properties, one
striking distinction remains. In Fig. \ref{fig:fig3} we present distributions of the bulge-to-total luminosity ratios.
Grand-design galaxies appear to have significantly more massive bulges than multi-armed galaxies, while sharing similar
bulge effective radii and S\'ersic indices (see Fig.~4 in \cite[Bittner et al. 2017]{bittner2017}). Thus, the main
difference we find in the fundamental galaxy properties of multi-armed and grand-design galaxies is the density of their
bulge component. 

A possible explanation that reconciles the similarities of their host discs with the striking differences in their 
spiral pattern is related to the theory of spiral wave modes. 
Lindblad (1959) and Lin \& Shu (1964) proposed that spiral structure might the caused by spiral density waves. However,
the density waves in this theory do have an inward group velocity (Toomre 1969) and thus would eventually wrap up,
unless reflected off a central component with a high Toomre-Q parameter. Such a reflection would result in a leading
wave with an outward group velocity.  Subsequently, this leading wave might be amplified at corotation by the swing
amplification mechanism (Toomre 1981), thus producing a strong trailing wave with an inward group velocity. In this framework of a positive feedback
loop a persistent, standing spiral wave mode can develop (Lin 1970; Mark 1976a,b,c, 1977; Bertin 1983; Lin \& Bertin
1985; Bertin et al. 1989a,b). In summary, the main structural condition for long-lasting spiral density waves is the
existence of a high Toomre-Q component in the centre of the galaxy. 

Such a high Toomre-Q region might be provided by the dense and massive bulges in grand-design galaxies. If this was
indeed the case, the spiral structure in grand-design galaxies would be caused by persistent spiral wave modes, which
would naturally explain the very high level of bi-symmetry of their spiral pattern.  On the contrary, multi-armed
galaxies would be governed by transient spiral structure characterised by a lower level of symmetry, as their less dense
bulges might simply not be able to reflect an incoming density wave.

In fact, these observational findings are consistent with recent numerical simulations.  \cite{saha2016} simulated a set
of individual galaxies in a sequence of various bulge configurations. They find that in galaxies with intermediate bulge
masses a long-lived, grand-design-like spiral pattern develops, well consistent with the theoretical framework and observational signatures
presented above. 

\end{document}